# Narrow-linewidth optical frequency comb reference to a fiber delay line

Haochen Tian,[a] Fei Meng,[b,c,d] Baike Lin[b], Shiying Cao,[b] Zhanjun Fang,[b] Youjian Song,[a,*] and Minglie Hu[a]

[a]Ultrafast Laser Laboratory, Key Laboratory of Opto-Electronic Information Technical Science of Ministry of Education, School of Precision Instrument and Opto-Electronics Engineering, Tianjin University, Tianjin 300072, China
[b]National Institute of Metrology, Beijing 100011, China
[c]Key Laboratory of Advanced Optical Communication System and Networks, School of Electronics Engineering and Computer Science, Peking University, Beijing 100871, China
[d]E-mail: mfei@nim.ac.cn

**Abstract**. In this letter, we derive a fully-stabilized narrow-linewidth optical frequency comb (OFC) reference to a kilometer-long fiber delay line for the first time, to the best of our knowledge. The 1537-nm comb modes and 1566-nm comb modes in the OFC are phase-locked to the fiber delay line with 40-kHz locking bandwidth. From out-of-loop measurement, the 1542-nm comb mode has residual phase noise of 925 mrad (integrated from 10 MHz to 1 kHz), fractional frequency stability of $9.13 \times 10^{-13}$ at 12.8 ms average time and 580 Hz linewidth. The linewidth has been compressed by a factor of ~ 170 compared to the free-running condition. Short-term stability of presented OFC exceeds most commercial microwave oscillators. The entire phase-locking system is compact and highly-integrated benefiting from absence of optical amplifiers, *f-2f* interferometers and optical/radio references. The presented OFC shows significant potential of being reliable laser source in low-noise-OFC-based precise metrology, microwave generation and dual-comb spectroscopic applications outside the laboratory.

## 1 Introduction

Optical frequency combs (OFCs) [1-4], which first emerged in 2000s, function as a bi-directional gear box to connect optical and radio world with a coherent manner, enabling phase link and comparison between microwave and optical frequency references [5-7], and between distinct optical frequency references [8-11]. Not surprisingly, over the past two decades, low-noise OFCs play essential roles in wide variety of high-precision applications such as distance ranging [12-14], dual-comb spectroscopy [15-17], to name only a few.

After the carrier-envelope offset frequency ($f_{ceo}$) and the repetition rate frequency ($f_{rep}$) are both phase-locked to an absolute radio frequency reference, e.g. a cesium clock, every optical mode in an OFC can be traced back in a phase coherent fashion to the absolute radio frequency reference. In this manner, atomic clocks are able to deliver its stability to every comb modes at $10^{-12} \sim 10^{-13}$ level in 1 s, resulting well-determined, equally-spaced comb teeth [18-19]. However, the linewidths of these OFCs are still not obviously narrowed down (at > kHz level) due to the inferior short-term stability of radio references. A practical approach to obtain narrow linewidth OFCs is to trace the comb modes to narrow-linewidth optical references, which is stabilized to high-finesse optical cavities using Pound-Drever-Hall (PDH) technique [20]. In this technique, $f_{beat}$ (obtained from the beat between an OFC's certain comb line and an external optical frequency reference) is phase-locked to a radio reference through fast cavity length feedback. Meanwhile $f_{ceo}$ is phase-locked to a radio reference. This method can push the comb modes' fractional stability down to $10^{-17}$ level (1 s average time), residual phase error to < 1 rad and linewidth to sub-mHz levels [11, 21-23]. In this way, narrow-linewidth OFCs with variety of emission bands and mode spacing up to ~ GHz have been demonstrated [24-26]. Using narrow-linewidth OFCs as accurate optical gears, comparison between optical atomic clocks with $10^{-18}$ fractional instability can be achieved [10]. Cavity-stabilized OFCs can also be applied in a reverse process to transfer the low-noise optical signals into microwave domain. By detecting one harmonic of the pulse repetition rate of the narrow-linewidth OFCs, an ultra-stable microwave with fractional frequency instability of $10^{-16}$ can be generated [27-28].

Aforementioned phase locking approaches require microwave or cavity-stabilized optical standards, making the whole system complicated and high-cost. On the other side, optical fibers are potential candidates of being reliable references for linewidth narrowing due to its superior short-term stability. Consequently, OFCs regarding a low-cost kilometer-long fiber delay line as reference has been presented in discussed literature [29-30]. In particular in Ref [30], Kwon et al reported a comb-noise stabilization method to phase lock individual comb mode of an OFC to the fiber delay line while $f_{ceo}$ to a radio reference. The fully-stabilized OFC has sub-$10^{-15}$-level fractional frequency stability and 28-Hz absolute linewidth. However, an *f-2f* self-referencing is still required, in which optical amplify and nonlinear broadening process can't be avoided. An alternative solution to totally get rid of direct $f_{ceo}$ stabilization is selecting two comb modes in an OFC and phase locking them to two optical references separately [31]. In this approach, all the comb modes in the OFC are capable to inherit stability of the optical reference. This type of phase-stabilized OFCs, benefiting from relative narrow linewidth, are widely serve as indispensable laser sources in dual-comb spectroscopic applications e.g. spectroscopy [31], distance measurements [32] and low-noise microwave generation [33].

With combination of two techniques mentioned above, we derive a compact, highly-integrated OFC comb mode noise suppression concept which regards a low-cost kilometer-long fiber delay line as reference using delayed self-heterodyne method. In this work, through high-speed pump power modulation and extra-cavity offset frequency modulation, 1537-nm comb modes and 1566-nm comb modes in the OFC are simultaneously phase-locked to a 1.25×2-km fiber delay line. From out-of-loop measurement, the 1542-nm comb mode has residual phase noise of 925 mrad (integrated from 10 MHz to 1 kHz), fractional frequency stability of $9.13\times10^{-13}$ at 12.8 ms average time and 580 Hz linewidth. The entire phase-locking system is compact and highly-integrated for the reason that it is free from optical amplifiers, *f-2f* interferometers, high-finesse cavities, optical or radio references. The presented narrow-linewidth OFC provides reliable laser source for low-noise-OFC-based precise metrology, microwave generation and dual-comb spectroscopic applications outside the laboratory.

## 2 Methods

### 2.1 OFC stabilization concept

Our OFC stabilization concept derives from delayed self-heterodyne (DSH) method, that was firstly implemented in the frequency noise detection and stabilization of CW lasers [34]. In DSH setups, kilometer-long fiber delay line is equivalent to a high-finesse cavity when serving as an external reference with a free spectral range (FSR) of ~ 400 kHz [35]. From the perspective of frequency domain, DSH technique is capable of phase-locking the optical frequency of target CW laser to one of the fiber delay line's transmission line ($v_{FSR}$). In order to derive this technique in optical frequency comb stabilization, control of two degrees of freedom, e.g. the repetition rate and carrier-envelope offset frequency is required. Through selecting two bunches of distinct comb modes with a certain gap in an OFC and phase locking them to the $v_{FSR}$ by certain feedbacks (repetition rate and carrier-envelope offset frequency control), frequency noise of all the optical modes in the OFC are controlled, as shown in Fig. 1. An out-of-loop measurement of beating a certain comb mode with a narrow linewidth CW laser could be conducted to evaluate the noise performance of the stabilized OFC.

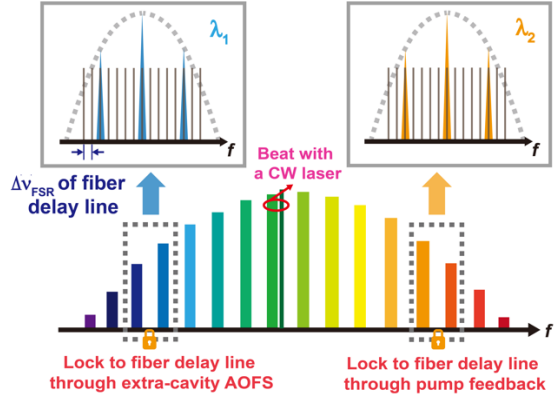

**Fig. 1.** Scheme of optical frequency comb reference to a fiber delay line

## 2.2 Laser source

The laser source used in our experiment is a home-made nonlinear polarization evolution (NPE) mode-locked Er-doped fiber laser with 205 MHz repetition rate. The output average power of the laser is ~120 mW (~0.5 nJ). The laser works at stretched pulse mode-locking regime with close to zero net-cavity dispersion. A fiber AOFS (AA Opto-Electronic, MT110-IIR20-Fio-SM0) is connected to the output of the laser to modulate $f_{ceo}$ of the laser at 25-MHz 3-dB bandwidth with transmission efficiency of 58%. On the other side, the 3-dB bandwidth for pump power modulation is 30 kHz, limited by the pump diode driver (Thorlabs, LDC 8020). A dense wavelength division multiplexer (DWDM) filters out 1537 nm, 1566 nm and 1542 nm for comb-line noise stabilization and out-of-loop comb mode stability evaluation, respectively. The filter bandwidth of DWDM is 200 GHz, corresponding to ~1.6 nm bandwidth in wavelength.

## 2.3 Asymmetric fiber interferometer

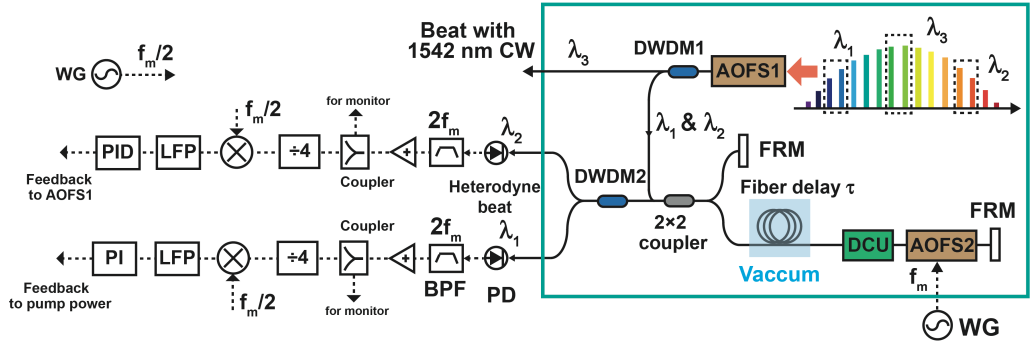

Fig. 2. Experimental setup. FRM, faraday rotating mirror; DWDM, dense wavelength division multiplexer; AOFS, acoustic optical frequency shifter; BPF, bandpass filter; LPF, lowpass filter; PD, photodetector (Menlo Systems, FPD510); PI: proportional-integral servo (Newfocus, LB1005); PID, proportional-integral-differential servo (Vescent Photonics, D2-125); DCU, delay control unit; WG, waveform generator.

An asymmetric fiber interferometer is implemented to generate the heterodyne beat for filtered wavelengths at $\lambda_1$=1537 nm and $\lambda_2$=1566 nm, as shown in Fig.2. A 2×2 fiber coupler splits the two filtered spectral segments into the reference arm and delayed arm. In the reference arm, two spectral segments are simply reflected back by an FRM at the end of the arm. In the delayed arm, two spectral segments pass through a 1.25-km long fiber spool (848-m SMF-28 + 402-m DCF38), a delay control unit (DCU), an AOFS and are reflected by an FRM at the end of the arm. The fiber spool is sealed by silicon rubber within a compact size (12-cm diameter) and put in a vaccum glassware. Such compact and robust fiber packaging enables the reduction of various technical noise induced by mechanical vibrations and temperature change at low Fourier frequency (<10 Hz). A waveform generator (Rigol, DG4162) drives the AOFS2, which is placed in the delayed arm shifting the optical signals' frequency twice at $2f_m$=100 MHz. Consequently, the pulses in long fiber arm has a frequency offset of 100-MHz-frequency ($2f_m$) from that in the reference arm. At the output of the asymmetric fiber interferometer, another DWDM de-multiplexes $\lambda_1$ and $\lambda_2$, which further illuminate two photodiodes for heterodyne beat detection.

Due to the extremely short coherent time of optical pulses from mode-locked lasers, one needs to slightly tune the laser's cavity length to guarantee that pulses of $\lambda_1$ overlap with its lateral ~2500th pulse temporally at the output of 2×2 coupler. This step is equivalent to overlapping the optical modes' frequencies around $\lambda_1$ with $\nu_{FSR}$ of the fiber delay line. To this end, 100-MHz heterodyne beat from $\lambda_1$ could be detected by the photodiode. To obtain heterodyne beat from $\lambda_2$, the delay in DCU needs to be carefully adjusted to compensate the fiber delay line's group velocity dispersion. Details on asymmetric fiber interferometer could be found in Ref [36].

The laser and asymmetric fiber interferometer are all put in an aluminum box and placed on a vibration isolation table (The Table Stable Ltd, AVI-200M/LP). The isolation table enables active control with cut-off frequency of 1 Hz. Through this careful package, acoustic noise induced by mechanical vibrations, air flow and temperature change could be partly rejected, leading to at least 6 hours continuous operation of narrow-linewidth OFC operation.

## 2.4 Phase-locking method

Two 100-MHz heterodyne beats from two wavelengths contain frequency noise information of filtered optical comb modes. We utilize two narrow bandpass filters (Telonic Berkeley, TTF72-5-5EE) with 100 MHz center frequency to filter out these two beats and amplify them to > 5 dBm. Two frequency dividers are used to divide 100-MHz heterodyne beats to 25 MHz, as shown in Fig. 2. Divided heterodyne beats are then mixed with $f_m/2$ from the same waveform generator. The discrimination results from the mixers are error signals with total frequency noise, $p \cdot [\tau(mf_{rep}+f_{ceo})]$, of filtered optical comb modes. Here, $m$ represents the mode number and $p$=980 represents the number of filtered comb modes. It should note that, between the amplifiers and frequency dividers, two radio power couplers split 1% of the heterodyne beats out for monitor. The error signal from 1566 nm is fed back upon the extracavity AOFS1 through a PID servo (Vescent Photonics, D2-125) for $f_{ceo}$ stabilization. On the other hand, the error signal from 1537 nm is fed back upon pump power modulation through a PI servo (Newfocus, LB1005) for repetition rate stabilization.

## 3 Results

### 3.1 Out-of-loop noise performance evaluation

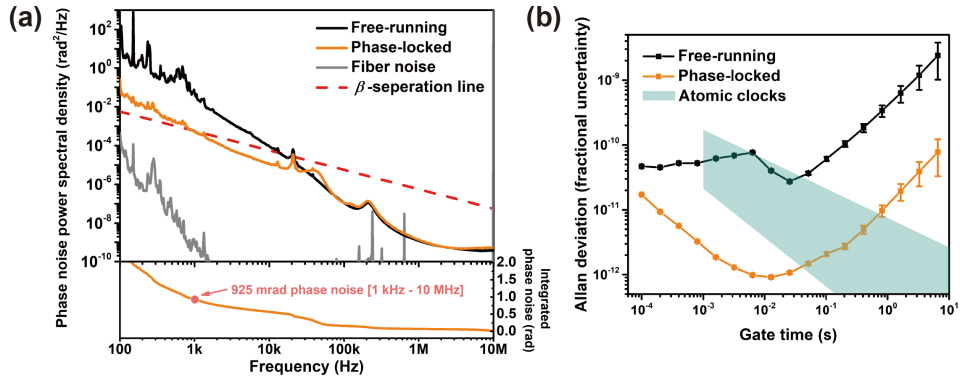

**Fig. 3.** Out-of-loop phase noise PSD (a) and frequency stability (b) of 1542-nm comb mode.

After close of two phase locked loops, two degrees of freedom of the target OFC are both stabilized, resulting a full-stabilized narrow linewidth OFC. The bandwidths of two phase locking loops are nearly 40 kHz, which is mainly limited by the first null frequency in the transfer function of fiber delay line (see Fig. S1. in supplementary material). To evaluate the noise performance of the stabilized frequency comb, comb mode around 1542 nm are filtered out by DWDM1 and beat with a commercial narrow-linewidth CW laser working at 1542 nm (Stable Laser System). The typical linewidth of the laser is 1 Hz. In this case, the phase noise, the Allan deviation and the linewidth of out-of-loop beat signal ($f_{beat}$) represent those of the individual comb mode at 1542 nm in the stabilized OFC because the reference CW laser has negligible noise compared to stabilized frequency comb.

The residual phase noise power spectral density (PSD) of $f_{beat}$ is characterized by a signal source analyzer (SSA) (Keysight, E5052b), as represented by orange curve in Fig. 3. Compared with the case when laser is free-running (the black curve), the phase noise within locking bandwidth is efficiently reduced. The resulting residual integrated phase noise is 925 mrad (integrated from 10 MHz to 1 kHz). The $\beta$-separation line (red-dashed curve in Fig. 3 (a)) indicates that estimated linewidth of 1542 nm comb mode is below 1 kHz. The noise of fiber is also measured and plotted in gray curve in Fig. 3. Detailed setup for fiber noise measurement could be found in section S2 in supplementary material. It should be noted that, at low Fourier frequency (< 100 Hz, not shown in Fig. 3), the phase noise PSD is mainly limited by the noise of fiber. This means that we have nearly fully used the capacity of phase-locked loops to suppress frequency noise at low Fourier frequency. The frequency variations of the $f_{beat}$ are recorded by a frequency counter (Keysight, 53220A) with 100 μs gate time. Fractional Allan deviation of $f_{beat}$ is calculated and presented in Fig. 3(b). The frequency stability has been improved by at least one order of magnitude after phase-locking. Allan deviation averages down as $\tau^{-1/2}$ until 10 ms and reaches $9.13\times10^{-13}$ at 12.8 ms. For longer time scale (> 12.8 ms), Allan deviation shows a drift behavior, which is affected by the long-term drifting of the referenced fiber delay line. Green shaded parts in Fig. 3(b) shows the shot-term stability of several atomic clocks. The comparison exhibits that OFC's short-term stability exceeds that of H masers at < 40 ms time scale, is superior than that of Cs clock at <100 ms time scale, is superior than that of Rb clock at < 1 s time scale. This reveals that fiber delay line stabilized OFC has comparative or even superior short-term stability than the OFCs that are traced to microwave references. Detailed data of comparison could be found in supplementary material.

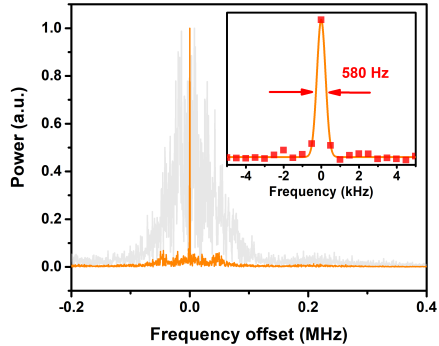

**Fig. 4.** Linewidth measurement of $f_{beat}$ before (grey curve) and after (orange curve) phase-locking. Inset: the linewidth is estimated to be 580 Hz from Gaussian fit.

Next step is to characterize the absolute linewidth of comb modes in the stabilized OFC. Through sampling the $f_{beat}$ at 500 MHz sampling rate in 2 ms using an oscilloscope (LeCroy WaveRunner, 104Xi), spectrum of $f_{beat}$ could be characterized through Fourier transformation, as shown in Fig. 4. Compared with the case when laser is free-running, the linewidth of the 1542-nm comb mode has been compressed from ~ 100 kHz (as represented by grey curve in Fig. 4) to 580 Hz (as represented by orange curve in Fig. 4) with a factor of ~ 170.

*3.2 Discussion*

From above out-of-loop measurements, we find that frequency stability is gradually degraded at > 10 ms average. In-loop phase noise PSD reaches those of the fiber delay line at low Fourier frequency (< 100 Hz) and drifts with it (See Section S3 in supplementary material). Consequently, it is reasonable to speculate that comb modes around 1537 nm and 1566 nm are drifting with the fiber delay line in long term. Thus long-term instability of the all the comb teeth in the OFC are degraded. Table 1 shows a comparison of representative narrow-linewidth OFCs. From Table 1, it is clear that OFCs referenced to fiber delay line have manifested two to three orders of magnitude improvement when compared with radio-frequency referenced OFCs. However, their linewidths are still not comparable to OFCs regarding cavity-stabilized lasers as reference. It is noteworthy that in the majority of applications, such as dual-comb spectroscopy and time-of-flight based distance measurements, linewidth have negligible impact on the measurement resolution when using ~ 100 Hz-linewidth OFCs as sources.

**Table 1** Linewidth comparison of representative narrow linewidth optical frequency combs with different reference.

| Literature | Stabilization scheme | Reference | Linewidth |
|---|---|---|---|
| [19] | $f_{rep}+f_{ceo}$ | RF references[a] | ~ 30 kHz |
| [22] | $f_{beat1}+f_{ceo}$ | Cavity-stabilized laser+RF reference | < 1 mHz |
| [11] | $f_{beat1}+f_{ceo}$ | Cavity-stabilized laser+RF reference | < 7.6 mHz |
| [25] | $f_{beat1}+f_{ceo}$ | Narrow-linewidth laser+RF reference | 3 kHz |
| [31] | $f_{beat1}+f_{beat2}$ | Cavity-stabilized lasers | 1 Hz |
| [30] | $f_{beat1}+f_{ceo}$ | Fiber delay line+RF reference | 28 Hz |
| This work | $f_{beat1}+f_{beat2}$ | Fiber delay line | 580 Hz |

[a]RF reference: radio-frequency reference.

The trend of development in OFCs is inclined to highly-integration, compactness, low-power consumption and practicality [37-38]. To summarize our work from this prospect, we explore a compact fully-stabilized narrow-linewidth OFC by phase-locking 1537-nm comb modes and 1566-nm comb modes to a 1.25×2-km fiber delay line. The out-of-loop measurement exhibits that 1542-nm comb mode has residual phase noise of 925 mrad (integrated from 10 MHz to 1 kHz), fractional frequency stability of $9.13\times10^{-13}$ at 12.8 ms average time and 580 Hz linewidth. Long-term frequency drifting of 1542-nm comb mode is mainly caused by the long-term instability of fiber delay line. More precise temperature feedback control to the whole optical system could probably overcome this issue. Even so, the short-term stability of the comb modes is still superior than microwave-referenced OFCs in short-term scale. Moreover, compare with state-of-the-art fully-stabilized OFC systems with *f-2f* interferometers, our phase-locking system gets rid of spectrum broadening process. Additionally, the entire phase-locking system is also compact and highly-integrated due to absence of external optical or radio references. Although a radio frequency source is needed, its stability is not specifically required because its noise is mixed out in DSH setup. This derived phase-locking method is not restrained for stabilization of mode-locked laser based OFCs, but could be also applied in comb mode noise suppression in EO-combs, QCL-combs, micro-combs, [39-40] etc. To prospect, our OFC is potential practical laser source in low-noise-OFC-based precise metrology, microwave generation and dual-comb spectroscopic applications (See design of relatively narrow-linewidth dual-comb system in supplementary material).

**Supplementary materials**

*S1  Transfer function of 1.25×2-km fiber delay line*

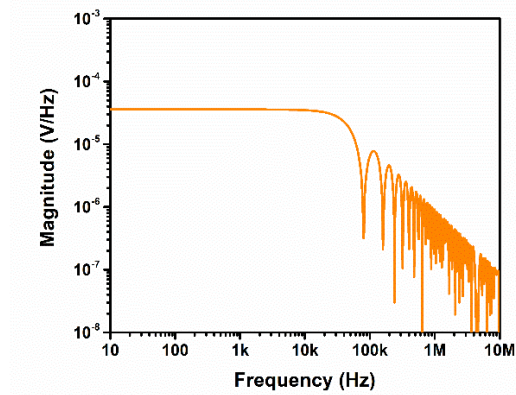

**Fig. S1** Transfer function of 1.25×2-km fiber delay line

Transfer function T($f$) of 1.25×2-km fiber delay line is calculated by: T($f$)=V$_{peak}$[(1-exp(-i×2π$\tau$))/(i×$f$)] [V/Hz], where $f$ is Fourier frequency, $\tau$ is the length of fiber delay line and V$_{peak}$ is the amplitude of the low-pass filtered mixer output voltage from interference pattern.

*S2  Setup for fiber delay line noise measurement*

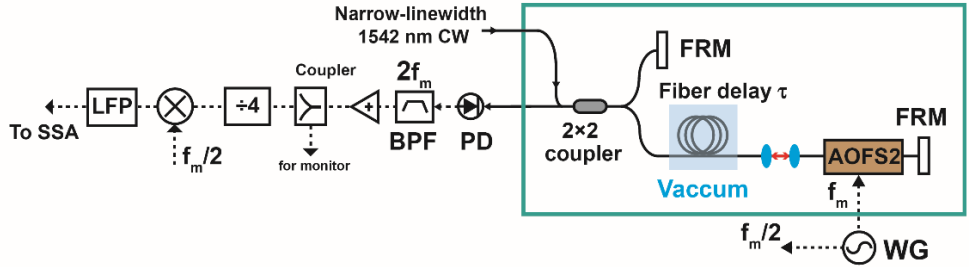

**Fig. S2** Experimental setup for fiber noise measurement. FRM, faraday rotating mirror; AOFS, acoustic optical frequency shifter; BPF, bandpass filter; LPF, lowpass filter; PD, photodetector; PI: proportional-integral servo; PID, proportional-integral-differential servo; WG, waveform generator.

Comb modes of $\lambda_1$ and $\lambda_2$ are replaced by a commercial narrow-linewidth CW laser at 1542 nm in fiber delay line noise measurement. The typical linewidth of the CW laser is 1 Hz, corresponding to a coherent length of 2×10$^5$ km in fiber. Thus, the heterodyne beat reflects the noise of a 1.25×2-km long fiber. To achieve this, the CW laser is directed into the same asymmetric fiber interferometer. Output heterodyne beat is detected, filtered, amplified, frequency-divided and finally mixed with $f_m$/2. Using the same data process introduced in Section 3.1, phase noise PSD introduced by the fiber delay line could be measured, as represented by grey curves in Fig. 3 and Fig. S4.

## S3  In-loop phase noise measurement

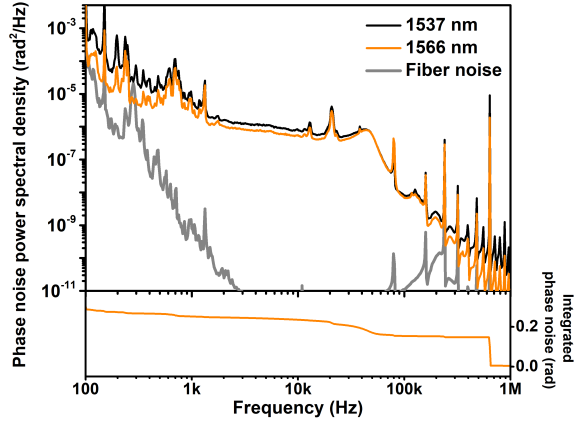

**Fig. S3.** In-loop residual phase noise PSD of 1537-nm and 1566-nm comb modes

The power spectral densities of in-loop voltage error signals from the monitor port of PI servo and PID servo could be characterized by SSA. Through division of the voltage PSD by fiber delay line's transfer function, in-loop residual phase noise PSDs of two wavelengths could be retrieved, as represented by orange curve and black curve in Fig. S3. The residual integrated phase noise is 302 mrad (integrated from 1 MHz to 100 Hz). At Fourier frequency above 40 kHz, PSD measurement is affected by null frequencies originated from fiber delay line.

## S4  Short-term stability comparison between atomic clocks

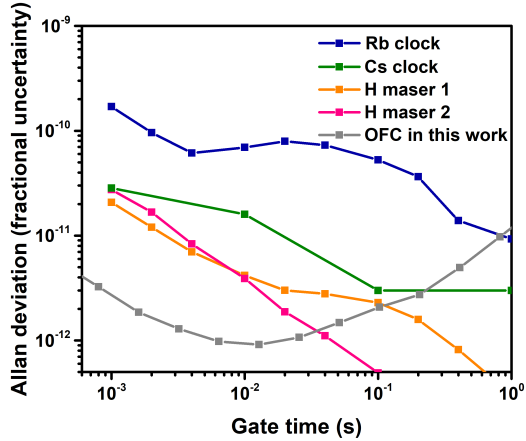

**Fig. S4.** Allan deviation of Rb clock (blue curve), hydrogen maser 1 (orange curve), hydrogen maser 2 (pink curve), Cs clock (green curve) and $f_{beat}$ (grey curve).

Fig. S4. shows the shot-term stability comparison between the presented OFC and several atomic clocks: a rubidium clock, a cesium clock and two hydrogen masers in our lab. The short-term stability is retrieved using state-of-the-art dual mixer time difference (DMTD) technique.

## S5 Design of relatively narrow-linewidth dual-comb system

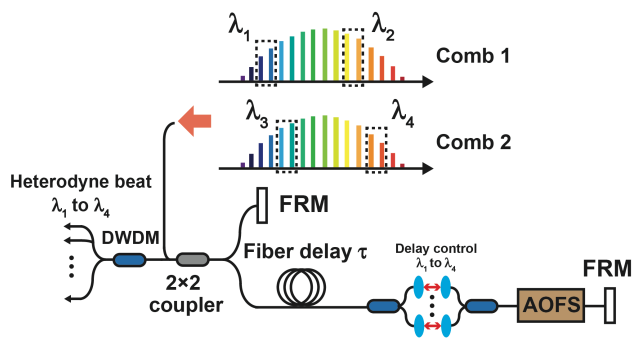

**Fig. S5.** Design of relatively narrow-linewidth dual-comb system. FRM, faraday rotating mirror; DWDM, dense wavelength division multiplexer; AOFS, acoustic optical frequency shifter.

As shown in Fig. S5 above, four bunches of comb modes of $\lambda_1$, $\lambda_2$, $\lambda_3$ and $\lambda_4$ from two separate lasers could be phase-locked to one same fiber delay line. Adjustment of the delay control unit in fiber interferometer could achieve dual combs with a certain repetition rate difference. The most intriguing point in this design is that common long-term drift between two lasers will be canceled due to the same fiber reference, resulting a relatively narrow-linewidth dual-comb system.

*Disclosures*

The authors declare no conflicts of interest.

*Acknowledgments*

The authors thank J. Kim and D. Kwon from KAIST for useful discussions.

**Haochen Tian** is a PhD student at Tianjin University.

**Caption List**

**Fig. 1** Scheme of optical frequency comb reference to a fiber delay line.

**Fig. 2** Experimental setup.

**Fig. 3** Out-of-loop phase noise power spectral density (a) and frequency stability (b) of 1542-nm comb mode.

**Fig. 4** Linewidth measurement of fbeat before (grey curve) and after (orange curve) phase-locking.